\documentclass[final,1p,times]{elsarticle} 
\usepackage{graphicx} 
\usepackage{amssymb} 
\usepackage{amsthm} 
\usepackage{lineno} 

\journal{Nuclear Physics A} 
\begin{document} 

\def\eq#1{{Eq.~(\ref{#1})}}
\def\fig#1{{Fig.~\ref{#1}}}
\newcommand{\as}{\alpha_s}

\begin{frontmatter} 


  \title{Early Time Dynamics in Heavy Ion Collisions from CGC and from
    AdS/CFT}

\author{Yuri V.\ Kovchegov}

\address{Department of Physics, The Ohio State University, Columbus,
  OH 43210, USA}

\begin{abstract} 
  We review two different theoretical approaches to the strong
  interaction dynamics at the early times immediately following heavy
  ion collisions. One approach is based on small-coupling physics of
  the Color Glass Condensate (CGC).  The other approach is based on
  Anti-de Sitter space/Conformal Field Theory (AdS/CFT) correspondence
  and may be applicable to describing large-coupling QCD interactions.
  We point out that in terms of theoretical tools the two approaches
  are somewhat similar: in CGC one deals with classical gluon fields
  produced in a nuclear shock wave collision, while in AdS/CFT one
  studies classical gravity in a gravitational shock wave collision.
  We stress, however, that the resulting physics is different: the
  classical gluon fields in CGC lead to a free-streaming medium
  produced in heavy ion collisions, while the classical gravity in the
  5-dimensional AdS bulk is likely to lead to ideal hydrodynamics
  description of the produced medium. Also, the valence quarks in
  colliding nuclei in CGC continue along their light cone trajectories
  after the collision with very little recoil, while we show that in
  AdS the colliding nuclei are likely to lose most of their energy in
  the collision and stop.
\end{abstract} 

\end{frontmatter} 




\section{Classical Gluon Fields in CGC}
\label{CGC}

We begin by considering a high energy heavy ion collision. The basic
premise of the CGC physics is that for each of the colliding
ultrarelativistic nuclei the small-$x$ wave function is characterized
by the large transverse momentum scale called the {saturation scale}
and denoted by $Q_s$. At high energy and/or for large nuclei the
saturation scale is large, $Q_s \gg \Lambda_{QCD}$, making the strong
coupling constant small, $\as \ll 1$, thus allowing for a
small-coupling description of the small-$x$ nuclear wave functions.
Collisions of two nuclei with such wave functions would lead to
interactions characterized by perturbatively large saturation scales
as well, allowing for a small-coupling description of the early stages
of heavy ion collisions. We refer the reader to
\cite{Jalilian-Marian:2005jf} for a review of CGC physics in nuclear
collisions.

If one is interested in the dynamics of the medium produced in the
collisions over a not very broad rapidity interval ($\Delta y \le
1/\as$), then the relevant collision dynamics is described in the
framework of the McLerran--Venugopalan (MV) model
\cite{McLerran:1993ni}. The MV model states that, due to the high
parton density in the colliding nuclei, the dominant gluon fields
produced in heavy ion collisions are classical, and are described by
the classical Yang-Mills equations
\begin{equation}\label{YM}
  D_\nu F^{\mu\nu} = J^\mu
\end{equation}
with the source current $J^\mu$ given by the color charges in the
colliding nuclei. 
\begin{figure}[th]
  \begin{center}
    \includegraphics[width=7cm]{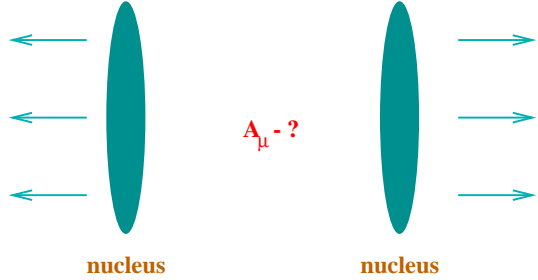}
  \end{center}
  \caption{Heavy ion collision in McLerran--Venugopalan model.}
  \label{AA1}
\end{figure}
This setup is illustrated in \fig{AA1}, where the two nuclei moving
away from each other provide the source current, and the classical
gluon field is left behind the two nuclei.

\begin{figure}[th]
  \begin{center}
    \includegraphics[width=7cm]{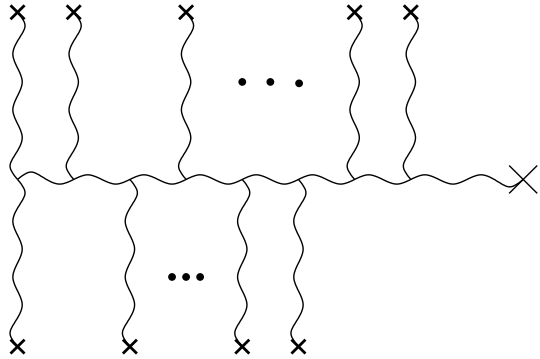}
  \end{center}
  \caption{Diagrammatic representation of the classical gluon field 
    produced in a heavy ion collision in McLerran--Venugopalan model.
    The thick crosses represent nucleons in the nuclei with the top row
    of the crosses denoting one nucleus and the bottom row denoting
    the other nucleus. The thin cross denotes the point in space-time
    where we measure the gluon field. Incidentally, the same diagram
    describes classical graviton field produced in a collision of two
    shock waves in AdS$_5$.}
  \label{AA2}
\end{figure}

The exact analytical solution of Yang-Mills equations (\ref{YM})
providing gluon field generated in a heavy ion collision does not
exist due to the complexity of the problem. In the diagrammatic
language to find the classical gluon field one has to resum an
infinite set of Feynman diagrams, and example of which is shown in
\fig{AA2}. There exist however perturbative solutions
\cite{Kovner:1995ts,Kovchegov:1997ke,Balitsky:2004rr}, an analytic
solution for the gluon production cross section in proton--nucleus
(pA) collisions \cite{Kovchegov:1998bi} and a numerical solution of
the full nucleus--nucleus (AA) problem \cite{Krasnitz:2002mn}. There
is also an analytical ansatz for the full solution for the gluon
production cross section in AA collisions \cite{Kovchegov:2000hz}.

\begin{figure}
  \begin{center}
    \includegraphics[width=7cm]{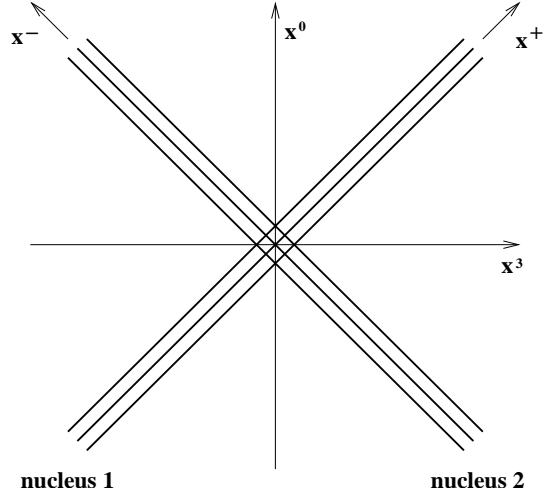}
  \end{center}
  \caption{The space-time picture of the ultrarelativistic heavy ion 
    collision in the center-of-mass frame. The collision axis is
    labeled $x^3$, the time is $x^0$.}
  \label{spacetime}
\end{figure}

To understand the matter produced by the classical gluon fields in AA
collisions let us first note that the distribution of this classical
matter is rapidity-independent. In the MV model the nuclei have a very
large transverse extent and are translationally invariant in the
transverse direction. The matter distribution thus does not depend on
the transverse coordinate $x_\perp$ and on the space-time rapidity
$\eta = (1/2) \ln [(x^0 + x^3)/(x^0 - x^3)]$, and depends only on the
proper time $\tau = \sqrt{(x^{0})^2 - (x^{3})^2}$. (Here $x^0$ is time
and $x^3$ is the collision axis: see \fig{spacetime} for the
explanation of the coordinates.) The most general energy-momentum
tensor of a matter distribution dependent only on $\tau$ can be shown
to be of the following form at mid-rapidity ($x^3 =0$)
\cite{Kovchegov:2005ss}
\begin{equation}\label{emt}
  T^{\mu\nu} \, = \,
  \left( \begin{array}{cccc} \epsilon (\tau) & 0 & 0 & 0 \\
      0 & p (\tau) & 0 & 0 \\
      0 & 0 & p (\tau) & 0  \\
      0 & 0 & 0 & p_3 (\tau) \end{array} \right)
\end{equation}
in the $x^0, x^1, x^2, x^3$ coordinates. In general the transverse
pressure $p (\tau)$ is not equal to the longitudinal pressure $p_3
(\tau)$.
Imposing the energy-momentum conservation condition $\partial_\mu
T^{\mu\nu} =0$ on \eq{emt} yields
\begin{equation}\label{eom}
  \frac{d \epsilon}{d \tau} \, = \, - \frac{\epsilon + p_3}{\tau}.
\end{equation}
Classical gluon field dynamics is conformal: therefore $T_\mu^\mu =0$
which implies $\epsilon = 2 p + p_3$.

At early proper times $\tau \ll 1/Q_s$ the classical gluon fields lead
to the following scaling of energy density $\epsilon (\tau)$
\cite{Lappi:2006hq,Fukushima:2007ja}
\begin{equation}\label{e_early}
  \epsilon (\tau) \bigg|_{\tau \ll 1/Q_s} \, \sim \, \ln^2
  \frac{1}{\tau \, Q_s}.
\end{equation}
With the help of \eq{eom} this leads to the following early-time
energy-momentum tensor of the produced matter
\begin{equation}\label{emt_early}
  T^{\mu\nu} \bigg|_{\tau \ll 1/Q_s} \, = \,
  \left( \begin{array}{cccc} \epsilon (\tau) & 0 & 0 & 0 \\
      0 &  \epsilon(\tau) & 0 & 0 \\
      0 & 0 & \epsilon (\tau) & 0  \\
      0 & 0 & 0 & - \epsilon (\tau) \end{array} \right) \, .
\end{equation}
Note that the longitudinal pressure at early times is negative $p_3 =
- \epsilon$. This result is true for any rapidity-independent medium
distribution at early time which has a finite total energy. 

At late proper times $\tau \gg 1/Q_s$ both the analytical perturbative
approaches \cite{Kovchegov:2005ss} and the full numerical simulations
\cite{Krasnitz:2002mn} lead to the energy density scaling as
\begin{equation}\label{e_late}
  \epsilon (\tau) \bigg|_{\tau \gg 1/Q_s} \, \sim \, \frac{1}{\tau}.
\end{equation}
This gives the following energy-momentum tensor
\begin{equation}\label{emt_late}
  T^{\mu\nu} \bigg|_{\tau \gg 1/Q_s} \, = \,
  \left( \begin{array}{cccc} \epsilon (\tau) & 0 & 0 & 0 \\
      0 &  \epsilon(\tau)/2 & 0 & 0 \\
      0 & 0 & \epsilon (\tau)/2 & 0  \\
      0 & 0 & 0 & 0 \end{array} \right) \, .
\end{equation}
This late-time energy-momentum tensor has zero longitudinal pressure
and corresponds to free-streaming of the produced medium. One can see
that by arguing that the net energy of the produced medium is $E \sim
\epsilon \tau$. Hence $E$ is constant for the energy density from
\eq{e_late}, which can be understood as being due to non-interacting
particles free-streaming away from the collision point.


\section{Classical Gravity in AdS$_5$}
\label{AdS}

The classical CGC picture of heavy ion collisions is self-consistent,
in the sense that it assumes that $\as \ll 1$ and then justifies the
assumption by generating a large momentum scale $Q_s$. However, it
lacks an essential ingredient needed to describe heavy ion collisions:
it does not lead to ideal hydrodynamics, which is known to describe
RHIC data on particle spectra and elliptic flow rather well
\cite{Heinz:2001xi,Teaney:2000cw}. In the rapidity-independent case
considered above, ideal hydrodynamics was first considered by Bjorken
\cite{Bjorken:1982qr}. It has 
\begin{equation}\label{emt_bj}
  T^{\mu\nu} \, = \,
  \left( \begin{array}{cccc} \epsilon (\tau) & 0 & 0 & 0 \\
      0 &  p (\tau) & 0 & 0 \\
      0 & 0 & p (\tau) & 0  \\
      0 & 0 & 0 & p (\tau) \end{array} \right).
\end{equation}
For the ideal gas equation of state $\epsilon = 3 p$ it has
\begin{equation}\label{e_bj}
  \epsilon (\tau) \, \sim \, \frac{1}{\tau^{4/3}}.
\end{equation}
Obtaining (ideal) hydrodynamics in a first-principles calculation for
heavy ion collisions is the important problem of thermalization and/or
isotropization of the produced medium. Below we discuss this problem
assuming that strong-coupling non-perturbative QCD effects are
responsible for the onset of the hydrodynamic behavior. Since it is
unknown how to consistently study QCD at strong coupling using
analytic methods we will instead use AdS/CFT correspondence, which
would allow us to study a distant cousin of QCD --- the ${\cal N} =4$
super-Yang-Mills (SYM) theory.

AdS/CFT correspondence conjectures that the dynamics of ${\cal N} =4$
$SU(N_c)$ SYM theory in four space-time dimensions is dual to the type
IIB superstring theory on AdS$_5 \times$S$_5$ \cite{Maldacena:1997re}.
In the limit of large number of colors $N_c$ and large 't Hooft
coupling $\lambda = g^2 N_c$ (with $g$ the gauge coupling constant)
such that $N_c \gg \lambda \gg 1$, AdS/CFT correspondence reduced to
the gauge-gravity duality: ${\cal N} =4$ SU($N_c$) SYM theory at $N_c
\gg \lambda \gg 1$ is dual to (weakly coupled) classical supergravity
in AdS$_5$. Hence the gauge dynamics at strong coupling, which
includes all-orders quantum effects, is equivalent to the classical
dynamics of supergravity. Instead of summing infinite classes of
Feynman diagrams in the gauge theory or using other non-perturbative
methods, one can simply study classical supergravity in 5 dimensions.
For a review of AdS/CFT correspondence see \cite{Aharony:1999ti}.

In a real-life heavy ion collision at RHIC the early-time dynamics is
likely dominated by the weak-coupling CGC physics. Strongly-coupled
dynamics may set in only at later times $\tau \sim 1/\Lambda_{QCD}$,
though even this time estimate is very crude. As can be shown using
the techniques of \cite{Kovchegov:2007pq}, matching of perturbative
CGC physics onto AdS/CFT dynamics at later times is not unique, and
does not lead to a single uniquely defined dual geometry in AdS$_5$,
allowing instead for a variety of possible metrics. Indeed the
expectation value of a single local operator (the energy momentum
tensor) coming from CGC can not uniquely constrain the full quantum
state of the field theory.  To keep our calculations under theoretical
control we will consider the whole collision of two nuclei in the
strong coupling AdS/CFT framework, understanding that this is simply a
rough approximation of the real heavy ion collision.

Our goal is to describe the isotropization (and thermalization) of the
medium created in heavy ion collisions assuming that the medium is
strongly coupled and using AdS/CFT correspondence to study its
dynamics. We want to construct a metric in AdS$_5$ which is dual to an
ultrarelativistic heavy ion collision as pictured in \fig{spacetime}.

We start with a metric for a single shock wave moving along a light
cone \cite{Janik:2005zt}:
\begin{equation}\label{1nuc}
  ds^2 \, = \, \frac{L^2}{z^2} \, \left\{ -2 \, dx^+ \, dx^- + \frac{2
      \, \pi^2}{N_c^2} \, \langle T_{--} (x^-) \rangle \, z^4 \, d
    x^{- \, 2} + d x_\perp^2 + d z^2 \right\}.
\end{equation}
Here $x^\pm = \frac{x^0 \pm x^3}{\sqrt{2}}$, $z$ is the coordinate
describing the 5th dimension such that the boundary of the AdS space
is at $z=0$, and $L$ is the curvature radius of the AdS space.
According to holographic renormalization \cite{deHaro:2000xn},
$\langle T_{--} (x^-) \rangle$ is the expectation value of the
energy-momentum tensor for a single ultrarelativistic nucleus moving
along the light-cone in $x^+$-direction in the gauge theory.

\begin{figure}[h]
  \begin{center}
    \includegraphics[width=3cm]{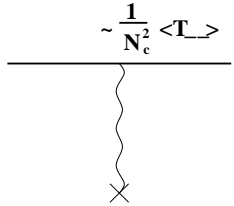}
  \end{center}
  \caption{A representation of the metric (\protect\ref{1nuc}) as a graviton 
    (wavy line) exchange between the nucleus at the boundary of AdS
    space (the solid line) and the point in the bulk where the metric
    is measured (denoted by a cross). }
  \label{1gr}
\end{figure}

The metric in \eq{1nuc} is an exact solution of Einstein equations in
AdS$_5$: $R_{\mu\nu} + \frac{4}{L^2} \, g_{\mu\nu} = 0$. It can also
be represented perturbatively as a single graviton exchange between
the source nucleus near the AdS boundary and the location in the bulk
where we measure the metric/graviton field. This is shown in
\fig{1gr}, where the solid line represents the nucleus and the wavy
line is the graviton propagator. Incidentally a single graviton
exchange, while being a first-order perturbation of the empty AdS
space, is also an exact solution of Einstein equations. This means
higher order tree-level graviton diagrams are zero (cf. classical
gluon field of a single nucleus in covariant gauge in the CGC
formalism \cite{Kovchegov:1997pc}).

\begin{figure}
  \begin{center}
    \includegraphics[width=14cm]{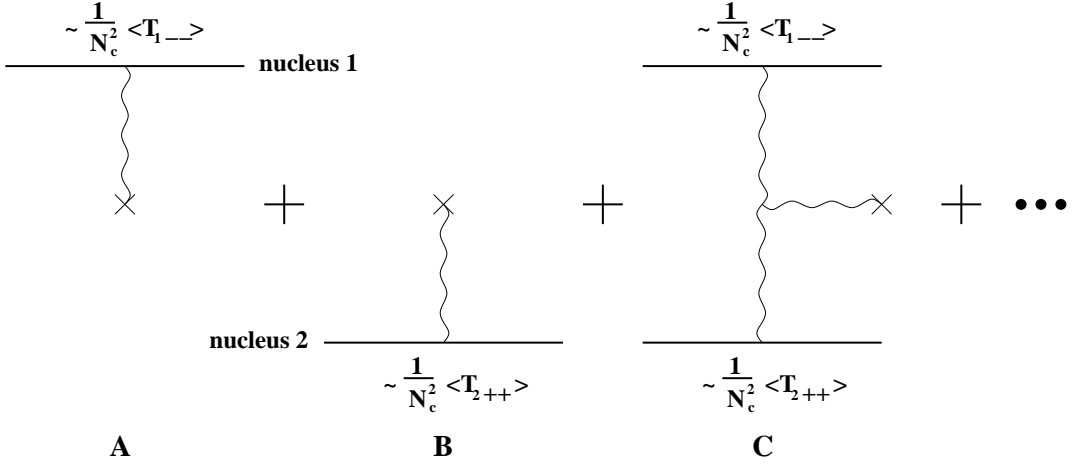}
  \end{center}
  \caption{Diagrammatic representation of the metric in \protect\eq{2nuc1}. 
    Wavy lines are graviton propagators between the boundary of the
    AdS space and the bulk.  Graphs A and B correspond to the metrics
    of the first and the second nucleus correspondingly.  Diagram C is
    an example of the higher order graviton exchange corrections.}
  \label{pert}
\end{figure}

Now let us try to find the geometry dual to a collision of two shock
waves with the metrics like that in \eq{1nuc}. We will follow
\cite{Albacete:2008vs}.  Defining $ t_1 (x^-) \, \equiv \, \frac{2 \,
  \pi^2}{N_c^2} \, \langle T_{1 \, --} (x^-) \rangle$ and $t_2 (x^+)
\, \equiv \, \frac{2 \, \pi^2}{N_c^2} \, \langle T_{2 \, ++} (x^+)
\rangle$ we write the metric resulting from such a collision as
\begin{eqnarray}\label{2nuc1}
  ds^2 \, = \, \frac{L^2}{z^2} \, \bigg\{ -2 \, dx^+ \, dx^- + d
  x_\perp^2 + d z^2 + t_1 (x^-) \, z^4 \, d x^{- \, 2}  
  + t_2 (x^+) \, z^4 \, d
  x^{+ \, 2} \nonumber \\ + \, \mbox{higher order graviton exchanges}
  \bigg\}
\end{eqnarray}
The metric of \eq{2nuc1} is illustrated in \fig{pert}. The first two
terms in \fig{pert} (diagrams A and B) correspond to one-graviton
exchanges which constitute the individual metrics of each of the
nuclei, as shown in \eq{1nuc}. We need to calculate the next order
correction to these terms, which is shown in the diagram C in
\fig{pert}.

\fig{pert} illustrates that construction of dual geometry to a shock
wave collision in AdS$_5$ consists of summing up all tree-level
graviton exchange diagrams. It is similar diagrammatically to the
classical gluon field formed by heavy ion collisions in CGC
\cite{Kovner:1995ts,Kovchegov:1997ke}. The diagram for the gluon field
shown in \fig{AA2} can also be interpreted as the diagram for
gravitons in AdS$_5$ bulk leading to the metric produced in a
collision of two gravitational shock waves.

Without going into details of the calculation which can be found in
\cite{Albacete:2008vs} one can calculate the graph in \fig{pert}C as
follows. Let us take delta-function shapes for the shock waves: $t_1
(x^-) \, = \, \mu_1 \, \delta (x^-)$ and $t_2 (x^+) \, = \, \mu_2 \,
\delta (x^+)$. The contribution of diagram C in \fig{pert} to the
energy density of the produced medium $\epsilon$ should therefore be
proportional to $\mu_1 \, \mu_2$. As $\mu_1$ and $\mu_2$ have
dimensions of mass cubed each ($M^3$), and dimension of $\epsilon$ is
$M^4$, we need to multiply $\mu_1 \, \mu_2$ by the square of proper
time, $\tau^2$, as this is the only dimensionful variable left. One
obtains \cite{Grumiller:2008va,Albacete:2008vs}
\begin{equation}\label{e_ads}
  \epsilon (\tau) \bigg|_{\tau \ll \frac{1}{\mu_1^{1/3}},
    \frac{1}{\mu_2^{1/3}}} \, \sim \, \mu_1 \, \mu_2 \, \tau^2.
\end{equation}
A more detailed calculation fixes the prefactor in \eq{e_ads}
\cite{Grumiller:2008va,Albacete:2008vs} and also specifies the region
of validity of this result. Energy density is rapidity-independent,
because of the cancellation of rapidity factors due to graviton
exchanges in \fig{pert}C. The energy-momentum tensor corresponding to
the energy density from \eq{e_ads} is
\begin{figure}
  \begin{center}
    \includegraphics[width=7cm]{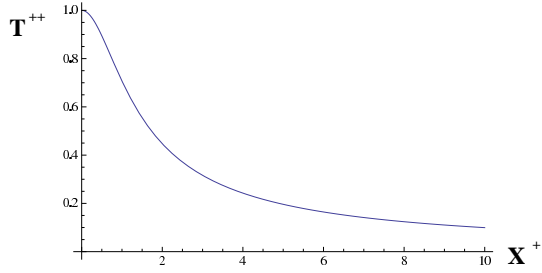}
  \end{center}
  \caption{$T^{++}$ component of the proton's energy-momentum tensor 
    after the collision as a function of light cone time $x^+$
    (arbitrary units).}
  \label{pstop}
\end{figure}
\begin{equation}\label{emt_ads}
  T^{\mu\nu} \, = \,
  \left( \begin{array}{cccc} \epsilon (\tau) & 0 & 0 & 0 \\
      0 & 2 \, \epsilon  (\tau) & 0 & 0 \\
      0 & 0 & 2 \, \epsilon (\tau) & 0  \\
      0 & 0 & 0 & - 3 \, \epsilon (\tau) \end{array} \right)
\end{equation}
and also has large negative longitudinal pressure, similar to the CGC
one (\ref{emt_early}). There is a problem with energy density scaling
as shown in \eq{e_ads} (or, equivalently, with the energy-momentum
tensor in \eq{emt_ads}): as was shown in \cite{Janik:2005zt}, in this
case there is a frame characterized by time 4-direction $t^\mu$ in
which the energy density is negative $T_{\mu\nu} \, t^\mu \, t^\nu <
0$. Hence the scaling of energy density in \eq{e_ads} with proper time
leads to negativity of energy density in other (boosted) frames. At
this point it is not clear whether this result presents a problem, as
there may be nothing wrong with energy density becoming negative for a
short period of time \cite{Grumiller:2008va}.

To better understand dynamics of the shock wave collisions let us
follow one of the shock waves after the interaction. First we
``smear'' the delta-function profile of that shock wave:
\begin{equation}\label{t1s}
  t_1 (x^-) \, = \, \frac{\mu_1}{a_1} \, \theta (x^-) \,
  \theta (a_1 - x^-).
\end{equation}
Here $\mu_1 \propto p^+ \, \Lambda^2 \, A^{1/3}$ and $a_1 \,
\propto \, R \, \frac{\Lambda}{p^+} \, \propto \,
\frac{A^{1/3}}{p^+}$, where the nucleus of radius $R$ has $A$
nucleons in it with $N_c^2$ valence gluons each. $p^{+}$ is the light
cone momentum of each nucleon and $\Lambda_1$ is the typical
transverse momentum scale.  The ``$+ +$'' component of the
energy-momentum tensor of a shock wave after the collision at $x^- =
a_1 /2$ is \cite{Albacete:2008vs}
\begin{equation}\label{stop}
  \langle T^{+ \, +} (x^+ , x^- = a_1 /2) \rangle \, = \, \frac{N_c^2}{2 \, \pi^2} \,
  \frac{\mu_1}{a_1} \left[ 1 - 2 \, \mu_2 \, x^{+\, 2} \, a_1 \right].
\end{equation}
The first term on the right of \eq{stop} is due to the original shock
wave while the second term describes energy loss due to graviton
emission. \eq{stop} shows that $\langle T^{+ \, +} \rangle$ of a
nucleus becomes {\sl zero} at light-cone times (as $\mu_2 \propto p^-
\, \Lambda^2 \, A^{1/3} \, \approx \, \mu_1$ in the center-of-mass
frame)
\begin{equation}\label{stoptime}
  x^+ \, \sim \, \frac{1}{\sqrt{\mu_2 \, a_1}} \, \sim \, 
\frac{1}{\Lambda \, A^{1/3}}. 
\end{equation}
Zero $\langle T^{+ \, +} \rangle$ would mean {\sl stopping} of the
shock wave and the corresponding nucleus. The result can be better
understood by doing all-order resummation of graviton exchanges with
one shock wave, which is needed for modeling proton-nucleus collisions
\cite{Albacete:2009ji}. The full result for the proton's ``$+ +$''
component of the energy-momentum tensor is
\begin{equation}\label{stop_pA}
  \langle T^{++} \rangle \, = \, \frac{N_c^2}{2 \, \pi^2} \,
  \frac{\mu_1}{a_1} \, \frac{1}{\sqrt{1 + 8 \, \mu_2 \, (x^+)^2 \,
      x^-}}, \ \ \ \mbox{for} \ \ \ 0 < x^- < a_1.
\end{equation}
\eq{stop_pA} is illustrated in \fig{pstop}, in which one can see that
the proton loses all of its light cone momentum over a rather short
time.

We thus conclude that the collision of two nuclei at strong coupling
leads to a necessary stopping of the two nuclei shortly after the
collision. If the nuclei stop completely in the collision, the strong
interactions between them are almost certain to thermalize the system,
probably leading to Landau hydrodynamics \cite{Landau:1953gs}. It is
possible that the mid-rapidity region of such collision may be
well-described by Bjorken hydrodynamics \cite{Bjorken:1982qr}, but
this still remains to be shown.


\section*{Acknowledgments}

This work is sponsored in part by the U.S. Department of Energy under
Grant No. DE-FG02-05ER41377.



\end{document}